\documentstyle[12pt]{article}

\begin{document}
\title{Exactly solvable random matrix models with additional two-body
interactions}
\author{K. A. Muttalib\\
Department of Physics, University of Florida\\
Gainesville, FL 32611.\\}
\maketitle
\begin{abstract}
It has been argued that despite remarkable success, existing random
matrix theories are not adequate to describe disordered conductors in
the metallic regime, due to the presence of certain two-body
interactions in the effective Hamiltonian for the eigenvalues, in
addition to the standard logarithmic interaction that arises entirely
from symmetry considerations. We present a new method that allows exact
solution of random matrix models with such additional two-body
interactions. This should broaden the scope of random matrix models in
general.\\
PACS Nos. 05.40, 72.10.Bg, 05.45.
\end{abstract}
\newpage
 From a phenomenological point of view, random matrix  models have
proved very useful in our understanding of a wide variety of physical
systems including complex nuclei \cite{Brody}, disordered metals
\cite{Stone} and chaotic systems \cite{Les}.
Although the physical systems are very diverse,  the local statistical
properties of
the characteristic levels of these systems in the bulk of the spectrum
turn out to be universal, similar to the
well-known universal properties of the distribution of eigenvalues
of random matrices as proposed originally by Wigner \cite{Mehta}. Recently
the models have been generalized to include
transitions in spectral statistics \cite{Muttalib} characteristic of
metal-insulator or
chaotic-regular transitions in finite systems. This has opened up the
possibility of describing such transitions in this powerful
mathematical framework, which allows exact evaluation of correlation
functions. However, on one hand the statistical properties of
numerically solved microscopic models with
random disorder describing mesoscopic conductors show remarkable
agreement with predictions of the generalized random matrix theory over a
wide range of disorder
\cite{Muttalib,Shlovski}; on the other hand there are
indications that the
appropriate random matrix model for disordered conductor is, while
highly accurate, not exact in the metallic regime
\cite{Slevin,Beenakker}. An exact solution \cite{Beenakker2} for
the Fokker-Planck equation describing the probability distribution of
the transmission coefficients \cite{Mello} shows that the resulting
matrix model
should include a small correction term which apparently destroys the
solvability of the model. This correction is responsible for a very
small correction to the magnitude of the universal conductance
fluctuation, but at the same time this also resolves a small discrepancy
between the random matrix result and the perturbative result from
microscopic theory \cite{Lee}. While it is not clear how important the
correction
term is to the question of e.g. transition from metal to insulator, the
fact that the correction  exactly reproduces the result of the
microscopic theory makes it qualitatively non-negligible. It is
therefore believed that despite remarkable success, the usefulness of
the phenomenological random matrix approach for the problem of
disordered conductors will be severely restricted, if such corrections
can not be accomodated within a solvable framework.
\par\indent
In this work we present a new method to accomodate certain type of
corrections to the standard random matrix models. These corrections are
similar to those arising in the problem of disordered conductors. The
method generalizes the
approach based on the theory of orthogonal polynomials, and
allows exact solutions for physically relevant models in terms of known
functions.
\par\indent
The basic ansatz of the random matrix theory is that for a physical
system described by an $N$x$N$ matrix $X$ with eigenvalues $x_n,
n=1,...N,$ the joint probabilty distribution (jpd) for
the ensemble of all random $X$  matrices consistent
with given symmetries (hermiticity, time reversal etc.) and subject to
some
physical constraint
(e.g. given average density of eigenvalues)
can be
written quite generally in the form \cite{Mehta}
\begin{equation}
P(x_1,.....x_N)=\prod_{m<n} |x_m-x_n|^{\alpha}\;\prod_n e^{-V(x_n)}.
\end{equation}
Here $\alpha$ is a symmetry parameter and is equal to 1, 2 or 4 for
orthogonal, unitary and symplectic symmetries respectively. For example for
disordered conductors, a good ansatz \cite{Muttalib2} is to use the $2N$x$2N$
matrix $X=\frac{1}{4}[TT^{\dag}+(TT^{\dag})^{-1}-2I],$ where $T$ is the
transfer matrix characterizing the conductor and $I$ is the unit matrix.
The doubly degenerate real eigenvalues $x$ are restricted between $0$ and
$\infty$ by current conservation, and directly gives the conductance
$g=\sum_{n}\frac{1}{1+x_n}.$ It is useful
to describe the probability distribution in
terms of an effective
``Hamiltonian'' $H$ of the
eigenvalues defined by $P=exp(-\alpha H)$, where
\begin{equation}
H(x_n)=-\sum_{m<n} ln|x_m-x_n|+\frac {1}{\alpha}V(x_n)
\end{equation}
The repulsive logarithmic `interaction' term arises from symmetry
considerations
alone, while the confining `single
particle potential' $V(x)$ is the Lagrange multiplier function which
takes care
of the physical constraint \cite{Balian} mentioned above, and in general
depends on
various physical
parameters. For example $V(x)=tx$, where $t$ depends on disorder, describes
the disordered metal quite well \cite{Stone,Chen}.
\par\indent
The solvability of the model has so far relied crucially on the
fact that the only interacting term in (2) is the logarithmic repulsion
which arises entirely from symmetry considerations; in other words any
relevant physical constraint must give rise to only a single particle
potential. Given this restriction, the universal distributions for
nearest neighbor spacing or the so called $\Delta_3$ statistics in the
bulk of the spectrum, which we will generically call the Wigner
distributions \cite{Mehta}, follow from the above jpd when $V(x)$ is taken
to be linear or quadratic
in $x$. In these cases the potential is strong enough to
overcome the logarithmic repulsion and the density of levels scale with
the number of levels. When $V(x)$ is not strong enough, the universality
breaks down; in particular for $V(x)\rightarrow [ln(x)]^2$ for large $x$
there
is a transition from the Wigner distribution to an uncorrelated Poisson
distribution as a function of a parameter \cite{Muttalib}. Nevertheless,
the model
remains exactly solvable.
\par\indent
The first hint that a disordered conductor in the metallic regime
in higher than one dimension may not be exactly described by a simple
logarithmic interaction of the above form came from attempts to check
detailed predictions of random matrix results numerically \cite{Slevin}.
But the
nature of the correction needed came from exact solutions
\cite{Beenakker2} of the
Fokker-Planck equation satisfied by the transmission eigenvalues in the
metallic regime \cite{Mello}. The solution showed, when mapped to a
random matrix
Hamiltonian, that the two particle interaction part has the form
\begin{equation}
-\frac {1}{2}\sum_{m<n} ln|x_m-x_n|-\frac
{1}{2}\sum_{m<n} ln|arcsinh^2(x_m^{1/2})-arcsinh^2(x_n^{1/2})|
\end{equation}
For $x<<1$, this reduces to the standard logarithmic repulsion,
but
for $x \approx 1$ the additional term makes it non logarithmic.
The difference is enough to change the variance of
conductance from the random matrix result $\frac{1}{8}$ to the
microscopic
perturbative result $\frac{2}{15}$. It is important to establish how
significant
this small difference is as far as the qualitative statistical
properties are concerned. But although the existence of such additional
two-body terms can be understood as arising from some physical
constraint that need not be of a single particle form, the question of
if or how it affects the known random matrix
results could not be addressed within the current random matrix
framework
because any such additional two-body interaction  destroys the
existing criterion for solvability, and therefore the usefulness, of the
model.
\par\indent
We will show below that with an additional two-body interaction given by
a simplification  of eq. (3), it is still possible to solve for the
model exactly, using a
new method. While the models constructed are
appropriate for disordered conductors, the solvability of such models
broadens the scope of
random matrix theory in  general.
\par\indent
As a first step towards constructing a model that can be solved exactly,
and is close to a physical model, we
approximate the $arcsinh$ function in (3) by a polynomial $s_k(x)$,
of degree $k$. In
the metallic regime where the model (3) is valid, this should be a good
approximation. For simplicity and purpose of illustration we will
discuss the case  $s_k(x)=x^k$ in detail. We will indicate at the end
how
the method can, in principle, be generalized to arbitrary polynomials.
\par\indent
We will therefore consider in detail the model described by an
additional two body interaction of the form $ln(x_m^k-x_n^k)$, which is
equivalent to a jpd of the form
\begin{equation}
P(x_1,.....x_N)=\prod_{m<n}
(x_m-x_n)(x_m^k-x_n^k)\;\prod_{n}e^{-V(x_n)}.
\end{equation}
where k is a positive integer.
Note that for k=1 the model reduces to the standard unitary random
matrix ensemble. An exact solution of this model should allow us to
understand at least the qualitative effects of the additional
two-body corrections.
\par\indent
For the standard logarithmic interaction part we follow the method of
orthogonal polynomials \cite{Mehta} and write the product of the
differences $\prod_{m<n} (x_m-x_n)$
as a Vandermonde determinant whose jth column has elements
$x^{j-1}_1,x^{j-1}_2,.....x^{j-1}_N$,
j varying from 1 to N. The determinant remains invariant if we add some
linear combination of the other columns with lower powers of the $x$'s;
the new jth column has elements
$Y_{j-1}(x_1),Y_{j-1}(x_2),.....Y_{j-1}(x_N)$, where
$Y_{j}(x)=\sum_{l=0}^{j} b_{jl}x^l$ is a polynomial in $x$, of degree
$j$; the coefficients $b$
will depend on the choice of the single particle potential $V$ as we
will show later. In a similar way, we write the correction term
$\prod_{m<n} (x^{k}_m-x^{k}_n)$ as a second Vandermonde determinant,
whose jth column has
elements $Z_{j-1}(x_1),Z_{j-1}(x_2),.....Z_{j-1}(x_N)$
where $Z_{j}(x)=\sum_{l=0}^{j} c_{jl}x^{kl}$ is now a polynomial in
$x^k$, of degree $j$; the coefficients
$c$ will be determined from the choice of $V$. We now multiply the ith
column
of each determinant by $exp[-V(x_i)/2]$, and interchange rows and
columns
of the second determinant. Eq. (4) can then be written as the product of
the two determinants, in the form
\begin{equation}
P(x_1,.....x_N)=C_N detK(x_i,x_j)
\end{equation}
where $C_N$ is the normalization constant, and
\begin{equation}
K(x_i,x_j)=
exp\left[-\frac{1}{2}\left(V(x_i)+V(x_j)\right)\right]
\sum_{l=0}^{N-1}Y_{l}(x_i)Z_{l}(x_j).
\end{equation}
The reason for writing the jpd as a determinant is the following. Our
ability to evaluate the  n-point correlation function, defined by
\begin{equation}
R_n(x_1,.....x_N)=\frac{N!}{(N-n)!}\int...\int dx_{n+1}...dx_N
P_{N}(x_1,.....x_N)
\end{equation}
depends on our ability to integrate the jpd over arbitrary number of
variables. These integrals can be done in a very simple way \cite{Mehta}
if the jpd
can be expressed as a determinant as in eq. (5), provided the following
two conditions are satisfied:
$$
(i) \int K(x,x)d\mu(x)=constant$$
\begin{equation}
(ii) \int K(x,y)K(y,z)d\mu(y)=K(x,z)
\end{equation}
where $d\mu$ is a suitable measure. This is where the
restriction of the standard logarithmic interaction, equivalent to the
case $k=1$, comes in. For $k=1$, the polynomial $Z(x)$ is the same as
$Y(x)$, and they can be chosen
such that they form an orthonormal set $p(x)$ with respect to the
measure
$exp[-V(x)]dx$, i.e.
\begin{equation}
\int p_m(x)p_n(x)e^{-V(x)}dx=\delta_{mn}.
\end{equation}
Then the above two conditions in (8) follow from the orthogonality and
normalizability of the polynomials. The additional two-body interaction
forces the polynomials to be distinct, destroying the orthogonality. We
will now show that even for
distinct polynomials $Y(x)$ and $Z(x)$, the above two conditions in (8)
can be satisfied
under some conditions, making it possible to obtain exact solutions for
the correlation functions for these generalized models.
\par\indent
Let us choose the coefficients $b$ and $c$ in such a way that the
polynomials $Y$ and $Z$ satisfy the following:
$$
\int Y_n(x)x^{kj}e^{-V(x)}dx=0,  j=0,1,....n-1,$$
$$
                                 \;\; \neq 0, j=n,$$
\begin{equation}
\int Z_n(x)x^{j}e^{-V(x)}dx=0,  j=0,1,....n-1,
\end{equation}
$$
                                \;\; \neq 0, j=n.$$
It can  then be shown \cite{Konhauser} that the two polynomials form a
``biorthogonal'' pair, defined by
\begin{equation}
\int e^{-V(x)}Y_n(x)Z_m(x)dx=g_n\delta_{mn}
\end{equation}
We will always choose an overall multiplicative factor such that
$g_n=1$,
i.e. the polynomials are normalized. Clearly the two conditions in (8)
are
then satisfied again,
$$
\int K(x,x)dx=\int e^{-V(x)}\sum_{l=0}^{N-1} Y_l(x)Z_l(x)dx$$
\begin{equation}
=\sum_{l=0}^{N-1}\int e^{-V(x)}Y_l(x)Z_l(x)dx=N
\end{equation}
where we have used the normalization, and
$$
\int K(x,y)K(y,z)dy=\int
e^{-\frac{1}{2}[V(x)+V(y)]}e^{-\frac{1}{2}[V(y)+V(z)]}
\sum_{j,l=0}^{N-1}Y_j(x)Z_j(y)Y_l(y)Z_l(z)dy$$
\begin{equation}
=e^{-\frac{1}{2}[V(x)+V(z)]}\sum_{j,l=0}^{N-1}Y_j(x)Z_l(z)\int
e^{-V(y)}Y_l(y)Z_j(y)dy
=K(x,z)
\end{equation}
where we have used the biorthogonality of the polynomials. Given these
properties, the integration over $N-n$ variables $x_{n+1},....x_N$ in
the jpd can be explicitly carried out \cite{Mehta}, and we
obtain
\begin{equation}
R_n=det[K(x_i,x_j)]_{i,j=1,...n}
\end{equation}
where the kernel $K(x_i,x_j)$ is given by eq. (6). In particular the
density is given by $K(x,x)$ and the two-particle kernel from which the
nearest neighbor spacing distribution or the $\Delta_3$ statistics can
be calculated is given by $K(x,y)K(y,x)$.
\par\indent
The model then is exactly solvable if for a given choice of the
single particle potential $V(x)$ the corresponding biorthogonal
polynomials can be obtained. The physically interesting model that
already gives a very good description
of the metallic regime of a disordered conductor in the $k=1$ limit is
given by $V(x)=x,
0 \leq x \leq \infty$. The model is exactly solvable in terms of Laguerre
polynomials. For arbitrary $k$, the corresponding biorthogonal
polynomials has been explicitly calculated by Konhauser
\cite{Konhauser2}. [For the
simplest non trivial case $k=2$, these are the polynomials introduced by
Spencer and Fano \cite{Spencer} to study penetration of matter by gamma
rays, and
studied later by Preiser \cite{Prieser}.] Therefore using the new method
the exact solution can be immediately written down in terms of these
Konhauser biorthogonal polynomials. It has been argued that an appropriate
generalization for all disorder, in the $k=1$ limit, is given by the
choice \cite{Chen}
\begin{equation}
V(x;q)=\sum_{n=0}^{\infty}\ln[1+(1-q)q^nx],\;\;\;\; 0\leq q<1.
\end{equation}
As $q\rightarrow 1^-, V(x)\rightarrow x,$ and one recovers the metallic
regime, while increasing
disorder corresponds to decreasing $q$. This model is exactly solvable
in terms of the $q$-Laguerre polynomials. For arbitrary $k$, again the
corresponding biorthogonal polynomials are explicitly known
\cite{AlSalam}, and  the exact solution can be written down in terms of
these ``q-Konhauser'' biorthogonal polynomials. The detailed properties
of these solutions are under investigation.
\par\indent
It is also possible to consider a more general form of the jpd
given by
\begin{equation}
P(x_1,.....x_N)=\prod_{m<n}
[r(x_m)-r(x_n)][(s(x_m)-s(x_n)\;\prod_{n}e^{-V(x_n)}.
\end{equation}
where $r(x)$ is a polynomial of degree $h$ and $s(x)$ is a polynomial of
degree $k$. Defining $Y$ and $Z$ as polynomials in $r(x)$ and $s(x)$
respectively,
the above method should be applicable if conditions (10) are replaced by
 \cite{Konhauser}
$$
\int e^{-V(x)}Y_n(x)[s(x)]^jdx=0,  j=0,1,....n-1,$$
$$
                                  \;\; \neq 0, j=n,$$
\begin{equation}
\int e^{-V(x)}Z_n(x)[r(x)]^jdx=0,  j=0,1,....n-1
\end{equation}
$$
                                \;\; \neq 0, j=n.$$
The case considered before is a special case where $r(x)=x$ and
$s(x)=x^k$.
Note that writing $r(x_n)-r(x_m)=(x_n-x_m)r(x_m,x_n)$ and
$s(x_n)-s(x_m)=(x_n-x_m)s(x_m,x_n)$, we can write the jpd in the form
\begin{equation}
P(x_1,.....x_N=\prod_{m<n}  (x_m-x_n)^2 \prod_{m<n}
r(x_m,x_n)s(x_m,x_n)\;\prod_{n}e^{-V(x_n)}
\end{equation}
which may allow more physically interesting models to be solved exactly,
if the corresponding biorthogonal polynomials are known.
\par\indent
In summary, we present a new method to accomodate certain two-body
interactions in random matrix models, particularly appropriate for
the problem of transport in disordered conductors. We show that
correlation functions can be written down explicitly in terms of known
biorthogonal polynomials. The approach should broaden the scope of
random matrix models in general.
\par\indent
I am grateful to Mourad Ismail for many discussions, and in particular
for bringing reference [19] to my attention.


\begin{thebibliography}{99}
\bibitem{Balian}R.Balian, Nuovo Cimento {\bf 57}, 183 (1968).
This is required
whenever the eigenvalues are un-bounded, while for
bounded eigenvalues
the confining potential is not necessary.
\end{thebibliography}
\end{document}